\documentstyle[epsf,epsfig,aps]{revtex}
\draft
\input epsf
\begin{document}
\twocolumn[\hsize\textwidth\columnwidth\hsize\csname@twocolumnfalse%
\endcsname
\title{Quantum Power Source. Putting in Order of a Brownian Motion without 
Maxwell's Demon
}
\author{V.V.Aristov and A.V.Nikulov}

\address{Institute of Microelectronics Technology and High Purity
Materials, Russian Academy of Sciences, 142432 Chernogolovka, Moscow
District, RUSSIA}

\maketitle
\begin{abstract}
{The problem of possible violation of the second law of thermodynamics is 
discussed. It is noted that the task of the well known challenge to the 
second law called Maxwell's demon is put in order a chaotic perpetual motion 
and if any ordered (non-chaotic) Brownian motion exists then the second law 
can be broken without this hypothetical intelligent entity. The postulate of 
absolute randomness of any Brownian motion saved the second law in the 
beginning of the 20th century when it was realized as perpetual motion. This 
postulate can be proved in the limits of classical mechanics but is not 
correct according to quantum mechanics. Moreover some enough known quantum 
phenomena, such as the persistent current at non-zero resistance, are an 
experimental evidence of the non-chaotic Brownian motion with non-zero 
average velocity. An experimental observation of a dc quantum power source 
is interpreted as evidence of violation of the second law.} \end{abstract}

\

 ] \narrowtext

\noindent
\textbf{1. INTRODUCTION.} 

The Brownian motion, the first mesoscopic phenomena, plays the important 
part in the history of physics. This phenomena was first observed as far 
back as two centuries ago. The investigations of the Brownian motion in the 
beginning of the 20 century shook the foundation of classical thermodynamics 
of the 19 century [1]. It was realized that it is the motion in the 
thermodynamic equilibrium state, i.e. the perpetual motion, which is not 
possible according to the old interpretation of the second law of 
thermodynamics predominant in the 19 century [2]. It ought be emphasized 
that the Brownian motion is experimental evidence not only of the perpetual 
motion but also of a perpetual driving force since no motion is possible 
without a driving force at non-zero friction. 

This driving force performs a work. But why can not we use this work? This 
problem is discussed already during more than century with essential benefit 
for science. The most known matter here is the Maxwell's demon. 

\bigskip

\noindent
\textbf{2. MAXWELL'S DEMON}

Maxwell's demon - a hypothetical intelligent entity capable of performing measurements on a thermodynamic system and using their outcomes to extract 
useful work - was considered a threat to the validity of the second law of thermodynamics for over a century [3]. It is no coincidence that this idea 
appeared at the same time with the Maxwell's kinetic theory of heat [4]. According to this theory the heat is the perpetual motion of atoms. Since 
absolute randomness of this motion was postulated one believed that the heat 
energy can be used for the performance of useful work only if it could be 
ordered even if partially. The partial regulating can be easy achieved under 
non-equilibrium conditions, for example at a temperature difference. But the 
task of the Maxwell's demon is to achieve the regulating under equilibrium 
conditions, when the total entropy might be systematically reduced, contrary 
to the second law of thermodynamics. Can exist the Maxwell's demon and if it 
can not exist then why? This problem has a long and interesting history 
which does not come to an end for the present. Because of the firm belief in 
the absolute status of the second law most scientists strove to exorcize the 
Maxwell's demon. Since the demon should obtain an information the process of 
the banishment is important not only for physics but also for the 
information theory [5]. 

The problem of the Maxwell's demon can be considered for a simple example of 
Szilard' engine [6]. Szilard considered in 1929 year a box that contains a 
single molecule, is capped at left and right ends by pistons, and is 
equipped with a movable partition which, when dropped, divides the box into 
equal left and right volumes. The molecule is maintained at temperature T by 
contact with the walls of the box. A cycle of the engine goes as follows: 
the partition, initially raised so that the molecule is free to explore the 
entire box, is dropped, and the demon determines an which side the molecule 
is trapped. Using this information, the demon inserts the piston on the 
empty side of the box, raises the partition, and allow the molecule to do 
isothermal work as it pushes the piston back to its original position. The 
demon extracts work $k_{B}T$, in apparent violation of the second law.

Different suggestions were proposed in order to save the second law. Brillouin assumed [7] that energy should be dissipated in observing the 
molecule's position. This point of view is developed up to last time [8]. 
Other way of the demon banishment, most popular in the last time [5,9-13], 
is the Landauer's principle. Landauer and others have found that almost any 
elementary information manipulation can in principle be done in a reversible 
manner, i.e. with no entropy cost at all [14]. Bennett [15] made explicit 
the relation between this result and the Maxwell's paradox by proposing that 
the demon can indeed learn where the molecule is in Szilard's engine without 
doing any work or increasing any entropy in the environment, and so obtain 
useful work during one stroke of the engine. But Bennett noted that an 
additional step is needed to complete the engine's cycle: the demon's memory 
stores one bit of information - molecule on right or left. To complete the 
cycle, this information must be erased as the demon's memory returns to a 
standard state, ready for the next cycle. Bennett invoked Landauer's 
principle -- to erase a bit of information in an environment at temperature 
T requires dissipation of energy $> k_{B}$\textit{Tln}2 -- and concluded that the demon 
does not succeed in turning heat into work. 

\bigskip

\noindent
\textbf{3. ORDERED BROWNIAN MOTION}

Although first doubts about the absolute status of the Landauer's principle 
were published already [16], most people believe that it saves the second 
law. But it should be noted that the Landauer's principle can save the 
second law only in the case of absolute randomness of any Brownian motion. 
Indeed, heat can not turn into work in Szilard's engine without the 
Maxwell's demon because of absolute randomness of the molecule's motion. The 
Maxwell's demon and also the ratchet/pawl combination considered by Feynman 
[17] (and earlier by Smoluchowski [1]) are needed in order to regulate 
chaotic heat energy. But if an ordered Brownian motion exists then the 
second law is broken without the Maxwell's demon and the ratchet/pawl 
combination. For example heat can be easy turned into work in Szilard's 
engine if the molecule moves in a direction with higher probability than in 
opposite direction. It is impossible in essence in the geometry considered 
by Szilard, but it is possible at a circular motion. For example in the case 
considered by Feynman [17] work can be obtained from heat without ratchet 
and pawl at an ordered circular motion of molecules. Therefore we can 
conclude that the postulate of absolute randomness of any Brownian motion 
saved the second law in the beginning of the 20th century when it was 
realized as perpetual motion. This postulate can be proved in the limits of 
classical mechanics but is not correct according to quantum mechanics.

According to the classical mechanics the average velocity of any Brownian motion equals zero $<v> = 0$ since if spectrum of permitted states is continuous then for any state with a velocity v a permitted state with opposite velocity $-v$ and the same probability $P(v^{2})$ exists, therefore $<v> = \Sigma _{per.st.} v P(v^{2}) + (-v) P(v^{2}) \equiv 0$. But according to the quantum mechanics no all states are permitted. Therefore the average velocity of some quantum Brownian motion can be non-zero $<v> \ne 0$. Thus, according to the well known principle of the quantum 
mechanics the postulate of absolute randomness of any Brownian motion can be incorrect. Moreover some enough known quantum phenomena are an experimental 
evidence of the non-chaotic Brownian motion with $<v> \ne 0$. 

\bigskip

\noindent
\textbf{3.1. Experimental evidence of non-chaotic Brownian motion}

One of the examples of the ordered Brownian motion is the persistent current observed at non-zero resistance [18]. The persistent current can exist because of the quantization of the momentum circulation 
$$\oint_{l} dlp  = \oint_{l} dl(mv + \frac{q}{c}A) = m\oint_{l} dlv +\frac{q}{c}\Phi = n2\pi \hbar  \eqno{(1)}$$
When the magnetic flux $\Phi $ contained within a loop is not divisible by the flux quantum $\Phi_{0} = 2 \pi \hbar c/q$ (i.e. $\Phi  \quad  \ne n\Phi_{0}$) and $\Phi  \ne  (n+0.5)\Phi _{0}$ the average velocity $<v>  \ne 0$ since the spectrum of permitted states of velocity circulation 
$$\oint_{l} {dlv}  = \frac{2\pi \hbar}{m}(n - \frac{\Phi}{\Phi 
_{0}}) \eqno{(1a)}$$
is discrete. Therefore the persistent current $j_{p} = qn_{q}<v>$, i.e. the direct current under equilibrium conditions, was observed at numerous experiments in superconducting [19] and even in normal metal [20-22] loops. First and most reliable experimental evidence of the persistent current at $R > 0$ is the Little-Parks experiment made first in 1962 year [23]. According to the universally recognized explanation [19] of this experiment the resistance oscillations $R(\Phi /\Phi_{0})$ are observed because of the oscillations of the persistent current $I_{p}(\Phi /\Phi_{0}) = sj_{p}(\Phi /\Phi_{0})$. The persistent current $I_{p}(\Phi /\Phi_{0} \propto  (<n> - \Phi /\Phi_{0})$ is a periodical function of the magnetic flux since the thermodynamic average value $<n>$ of the quantum number n is close to an 
integer number n corresponding to minimum energy, i.e. to minimum $(n - \Phi /\Phi_{0})^{2}$. Thus, according to the Little-Parks experiment and in spite of the Ohm's law $RI = -(1/c)d\Phi /dt$ a direct screening current flows along the loop [24] at a constant magnetic flux $\Phi  \ne n\Phi_{0}$ and $\Phi  \ne  (n+0.5)\Phi _{0}$ , i.e. without Faraday's voltage $-(1/c)d\Phi /dt = 0$.

\bigskip

\noindent
\textbf{3.2. Persistent current and Nyquist's noise.} 

The nearest classical phenomenon analogous to the persistent current at $R > 0$ is the Nyquist's (or Johnson's) noise. It is well known that any resistance at nonzero temperature is the power source of the thermally induced voltage [17]. This type of the Brownian motion was described theoretically by Nyquist [25] and was observed by Johnson [26] as long ago as 1928 year. Johnson observed a random voltage $<V^{2}> = 4R k_{B}T\Delta \omega $ in a frequency band $\Delta \omega $ on a resistance $R$ at a temperature $T$. Nyquist has shown that this voltage is induced by thermal fluctuation. It has the same value in frequency region from zero $\omega  = 0$ to the quantum limit $\omega  = k_{B}T/\hbar $. The observation of the persistent current at $R > 0$ as well as of the Nyquist's noise means that energy dissipation takes place: $RI_{p}^{2}$ in the first case and $<V^{2}>/R$ in the second case. Because both have power induced by fluctuations, the maximum power of the persistent current $RI_{p}^{2}$ [18] and to the total power of the Nyquist's noise are close to the power of thermal fluctuations $W_{fl} = (k_{B}T)^{2}/\hbar $. But there is an important difference between these two fluctuation phenomena. The power of the Nyquist's noise is "spread" $W_{Nyq} = k_{B}T\Delta \omega $ on frequency region from zero $\omega  = 0$ to the quantum limit $\omega  = k_{B}T/\hbar $ whereas the power of the persistent current is not zero at the zero frequency band $\omega  = 0$. It is very important difference. The persistent current can be interpreted as rectified Nyquist's noise. The Nyquist's noise is chaotic Brownian motion [17] and the persistent current at $R > 0$ is ordered Brownian motion [18]. Therefore the power of the first can not be used whereas the power of the second can be used for the performance of useful work.

\bigskip

\noindent
\textbf{3.3. Quantum force}

In order to describe the motion of Brownian particles Langevin has 
introduced a force which is called now Langevin force $F_{Lan}$. According to the Langevin equation 
$$m\frac{dv}{dt} + \gamma v = F_{Lan} \eqno{(2)}$$
we observe the Brownian motion of small particles in spite of non-zero friction $\gamma   \ne 0$ because of a random force $F_{Lan}$. According to the Langevin 
equation for the Nyquist's noise 
$$L\frac{dI_{Nyq} }{dt} + RI_{Nyq} =  E_{Lan}  \eqno{(3)}$$ 
a random current $I_{Nyq}$ flows along a loop because of a random fluctuation voltage $E_{Lan}$. Since the persistent current, as well as the Nyquist's one, is observed at non-zero resistance it is needed to introduce a force in order to explain why it is not damped. Such force was introduced in [18] and was called quantum force. 

According to [18] the persistent current is maintained in spite of the energy dissipation $RI_{p}^{2}$ because of reiterated switching of the loop between superconducting state with different connectivity induced by thermal fluctuation. When the superconducting state is unclosed the velocity of superconducting pairs is zero and the momentum circulation $\oint_{l} dlp  = \oint_{l} dl(mv + \frac{2e}{c}A) = m\oint_{l} dlv +\frac{2e}{c}\Phi =\frac{2e}{c}\Phi $ (see (1)). When the superconducting state is closed $\oint_{l} dlp  = n2\pi \hbar $ and the velocity can not be equal zero because of the quantization if $\Phi   \ne  n\Phi_{0} = n2\pi \hbar c/2e$. Therefore each superconducting pair accelerates and its momentum circulation changes from $(2e/c)\Phi $ to $n2\pi \hbar $ at each closing of superconducting state. This acceleration may be considered as an outcome of action of the Langevin force when the closing is induced by thermal fluctuation (as it takes place at the Little-Parks experiment). The change $(n2\pi \hbar  - (2e/c)\Phi) $ of the momentum circulation replaces random fluctuation voltage $E_{Lan}$. The velocity slows down and the momentum circulation returns to the initial value $(2e/c)\Phi $ because of dissipation force acting in the unclosed superconducting state when $R > 0$ as well as the Nyquist's current slows down at $E_{Lan} = 0$ because of $R > 0$ (see (3)). 

The Nyquist's noise is chaotic Brownian motion and the persistent current at $R > 0$ is ordered Brownian motion since the time average value of the Langevin force in the first case equals zero $< E_{Lan}> = 0$ whereas in the second case $(<n>2\pi \hbar - (2e/c)\Phi )\omega  = 2\pi \hbar (<n> - \Phi /\Phi_{0})\omega  \ne 0$ at $\Phi  \ne n\Phi_{0}$ and $\Phi  \ne  (n+0.5)\Phi _{0}$. The latter takes place because of discrete spectrum of closed superconducting state. Although the switching of the loop between superconducting state with different connectivity induced by thermal fluctuation is random (the frequency of switching $\omega  = N_{sw}/\Theta $, where $N_{sw}$ is a number of switching during a time $\Theta $) the quantum number $n$ at each closing has with high probability the same integer number $n$ corresponding to minimum energy i.e. to minimum $(n -\Phi /\Phi_{0})^{2}$, since the energy difference between adjacent permitted states with different n of superconducting loop is much higher than temperature. Therefore the average value $<n>$ is close to an integer number $n$ corresponding to minimum $(n -\Phi /\Phi_{0})^{2}$ and the quantum force 
$$\oint_{l} dlF_{q}  = 2\pi \hbar ( < n > - \frac{\Phi}{\Phi _{0}})\omega \eqno{(4)}$$
as well as the persistent current are a periodical function of the magnetic flux $\Phi $ inside the loop. The quantum force takes the place of the Faraday's voltage and maintains the persistent current in spite of the energy dissipation $RI_{p}^{2}$. 

\bigskip

\noindent
\textbf{4. NANO-SCALE QUANTUM POWER SOURCE.}

It is obvious that work can be easy obtained at an ordered circular motion of molecules, for example in the case considered by Feynman [17]. But how can we use the energy of the persistent current? It is doubtful that a work can be obtained at using of homogeneous, symmetric loop in which can not be 
a potential difference even at a non-zero current. But it is well known that a potential difference 
$$V = (<\rho >_{ls} - <\rho >_{l}) l_{s} j \eqno{(5)} $$
should be observed on a segment $l_{s}$ of an inhomogeneous conventional loop at a current density $j$ along the loop induced by the Faraday's voltage $j<\rho >_{l}l = <E>_{l}l = -(1/c)d\Phi /dt$ if the average resistivity along the segment $< \rho > _{l_{s}}  = \int_{l_{s}}  dl\rho / l_{s}$ differs from the one along the loop $ < \rho > _{l} = \oint_{l} dl\rho / l $. The relation (5) can be deduced from the Ohm' law $j\rho  = E = -\nabla V -- (1/c)dA/dt = -\nabla V - (1/cl)d\Phi /dt$. 

\bigskip

\noindent
\textbf{4.1. Persistent voltage.} 

If the persistent current $j_{p}(\Phi /\Phi_{0})$ is similar to the conventional current induced by the Faraday's voltage the persistent potential difference $V_{p}(\Phi /\Phi_{0}) = (<\rho >_{ls} - <\rho >_{l}) l_{s} j_{p}(\Phi /\Phi_{0})$ should be observed without an external current on segments of a inhomogeneous loop where $<\rho >_{ls} - <\rho >_{l} \ne 0$ and should not observed on segments of a homogeneous one where$<\rho >_{ls} - <\rho >_{l} = 0$. The experimental investigations [27] corroborate this analogy. 

The dependencies of the dc voltage $V$ on the magnetic flux $\Phi   \approx  BS$ of some round Al loops with a diameter $2r = $ 1, 2 and 4 $\mu m$ and a line width $w$ = 0.2 and 0.4 $\mu $m at the dc measuring current $I_{m}$ and different temperature close to $T_{c}$. The sheet resistance of the loops was equal approximately 0.5 $\Omega /\diamond $ at 4.2 K, the resistance ratio $R(300 K)/R(4.2 K)  \approx  2$ and the midpoint of the superconducting resistive transition $T_{c} \approx  1.24 \ K$. All loops exhibited the anomalous features of the resistive dependencies on temperature and magnetic field which was before observed on mesoscopic Al structures in some works [24,28]. 

In order to verify the analogy with a conventional loop both symmetric and asymmetric loops were investigated. Because of the additional potential contacts different segments of asymmetric loops have a different resistance at $T  \approx  T_{c}$ when $\Phi   \ne n\Phi_{0}$, whereas both segments of symmetric loops should have the same resistance if any accidental heterogeneity is absent. The conventional Little-Parks oscillations of the resistance were observed at the symmetrical loops. This result repeats the observations made before in many works and is not new result. In accordance with the analogy with a conventional loop (5) the voltage measured at contacts of symmetric loops 
equals zero at zero measuring current $I_{m} = 0$. In contrast to symmetrical loops no resistance but voltage oscillations $V(\Phi /\Phi_{0})$ proportional to the oscillations of the persistent current $j_{p}(\Phi /\Phi_{0})$ are observed on segments of asymmetric loop. In accordance with the prediction [18] and the analogy with a conventional loop (5) the voltage oscillations are observed without an external current.

The phenomenon observed in [27] was predicted first in [29]. It was shown first in this work that the dc voltage the value and sign of which depend in a periodic way on the magnetic flux can be observed on a segment of superconducting loop which is switched between normal and superconducting states. The value of this voltage should be proportional to the average frequency of the switching $\omega $, as well as the quantum force (see (4)), until the frequency does not exceed a limit one corresponded to a time relaxation. The analogy with a conventional loop (5) is conformed and the dc potential difference is observed [27] since the quantum force (4) as well as the Faraday's voltage $-(1/c)d\Phi /dt$ can not be localized in any segment of the loop in principle because of the uncertainty relation $\Delta p\Delta l > \hbar $ [18]. The velocity of superconducting pairs becomes nonzero when the momentum takes a certain value $\Delta p < p_{n + 1} -- p_{n} = 2\pi \hbar /l$, i.e. when superconducting pairs cannot be localized in any segment of the loop. The quantum force should be uniform along the loop: $\oint_{l} dlF_{q}  = lF_{q} $ .

From this relation and the relation (4) the connection between the frequency $\omega  = N_{sw}/\Theta $ of the switching between superconducting state with different connectivity and the voltage which should be observed on the loop segment remained all time in superconducting state can be deduced. Since the dissipation force does not act on superconducting pairs the balance of forces is $2eE = 2eV/l_{s} = F_{q} = 2\pi \hbar (<n> - \Phi /\Phi_{0})\omega /l$. Consequently the potential difference
$$V = \frac{\pi \hbar \omega}{e}( < n > - \frac{\Phi}{\Phi _{0}})\frac{l_{s}}{l} \eqno{(6)} $$ 
should be observed on a segment $l_{s}$ remaining in superconducting state when other segment is switched in normal state with frequency $\omega $. This relation between voltage and frequency resembles the Josephson one (see for example [30] ). The total balance of force circulation $\oint_{l} dlF_{q}  + \oint_{l} dl F_{dis} = 0$ explains why the persistent current is observed in spite of the dissipation $F_{dis} \ne 0$. This balance arises from the conditions that the total change of the momentum circulation during a long time should equal zero and that $\oint_{l} dl\nabla V \equiv 0$. The momentum circulation of superconducting pair changes from $(2e/c)\Phi $ to $n2\pi \hbar $ because of quantization and from  $n2\pi \hbar $ to $(2e/c)\Phi $ because of the dissipation force. During a time unity $(<n>2\pi \hbar - (2e/c)\Phi )\omega  + ((2e/c) \Phi  - <n>2\pi \hbar)\omega  = 2\pi \hbar (<n> - \Phi /\Phi_{0})\omega  + \oint_{l} dl F_{dis} = 0$. 

\bigskip

\noindent
\textbf{4.2. Persistent power.}

The observation of the voltage oscillations [27] is experimental evidence that the quantum force as well as the Faraday's voltage is distributed uniformly among the loop. This likeness between the quantum force and the Faraday's voltage explains why the analogy between the persistent current and the conventional current is corroborated [27]. But there is an important difference between these currents. The conventional current in accordance with the Ohm's law $j\rho  = E = -\nabla V -- (1/c)dA/dt = -\nabla V - (1/cl)d\Phi /dt$ has the same direction with the electric field in the whole of loop whereas the persistent current is observed without the Faraday's voltage $dA/dt = (1/l)d\Phi /dt = 0$ and consequently the electric field $E = -\nabla V$ and the persistent current $I_{p}$ should have opposite directions in a segment because $\oint_{l} dl\nabla V \equiv 0$. This means that according to the predictions [18,29] and the experimental result [27] a segment of the asymmetric loop is a dc power source:$VI_{p} \ne  0$ when $\Phi  \ne n\Phi_{0}$ and $\Phi  \ne  (n+0.5)\Phi _{0}$. It should be noted that already the classical Little-Parks experiment is evidence of the dc power source since the power dissipation $RI_{p}^{2}$ can be observed only if a power source $RI_{p}^{2}$ exists.

\bigskip

\noindent
\textbf{4.3. Direct-current generator}

Thus the theoretical [18,29] and experimental [27] investigations show that inhomogeneous mesoscopic superconducting loop can be used as direct-current generator the persistent power of which is induced by thermal fluctuations. Although the power of fluctuations is weak ($W_{fl} = (k_{B}T)^{2}/\hbar  \approx  10^{ - 8} \ Wt$ at $T$ = 100 K) enough power acceptable for real applications can be obtained since the power of the dc power source can be added. It is the second important difference of the persistent current from the Nyquist's noise. The power of the Nyquist's noise $W_{Nyq} = k_{B}T\Delta \omega $ observed on one resistance equals the one observed on $N$ resistance whereas the power of any $N$ dc power source can be added. Since the segment of the inhomogeneous loop is a dc power source the voltage $V_{N} = NV$ should be observed on a system of identical inhomogeneous loops segments of which are connected in series. The power $W_{load} = N^{2}V^{2}R_{load}/(R_{load}+NR_{s})^{2} = NV^{2}/4R_{s}$ can be obtained on an electric device with the resistance $R_{load} = NR_{s} $ loaded on this system [31]. Here $R_{s}$ is the resistance of the segment which is a load in the inhomogeneous loop. The persistent power observed on one inhomogeneous loop $W_{p,1} = V_{p}^{2}/R_{s} < I_{p}^{2}R_{l} $ can not exceed $(k_{B}T)^{2}/\hbar $ [18,29,32] because it is induced by thermal fluctuations. But the system of $N$ identical inhomogeneous loops $W_{p,N} = NV_{p}^{2}/R_{s} < N(k_{B}T)^{2}/\hbar $ can be enough powerful when the number of the loops $N$ is enough great. The $W_{p,N}/4$ part of this power can be used in an useful electric device. Such system can be used simultaneously as direct-current generator [33] and refrigerator [34]. 

Since the fluctuation power is proportional to $T^{2}$ it is better to use high-Tc superconductor (HTSC) with critical temperature $T_{c} \approx $ 100 K  for the making of the quantum power source on base of non-chaotic Brownian motion [31,35]. Since the value of effects connected with the persistent current in loops is proportional to $(l/\xi )^{2}$ and the coherence length $\xi $ of HTSC known now is small the loops should be nano-scale. The modern methods of nano-technology allow to make the system of $10^{8}$ loops on an area $ \approx  1 \ cm^{2}$. Such system of HTSC loops can give the dc power up to $W_{p,N} < N(k_{B}T)^{2}/\hbar  \approx  1 \ Wt$. The power can be increased in many times by the use of multi-layer technology. The power up to 10 kWt can be obtained in a system with volume $ \approx  100 \ cm^{3}$ and thickness of layers 0.01 cm.

Thus nano-scale quantum power source with acceptable power and acceptable volume can be made. But very high technology requires in order to make it.

\bigskip

\noindent
\textbf{5. DISCUSSION} 

It is obvious that the result [27] is experimental evidence of dc power source and that dc power can be used for the performance of useful work in contrast to the chaotic Nyquist's noise. But defenders of the second law do not retreat. They state that the dc power observed in [27] is induced by an external non-equilibrium electrical noise. Indeed, reiterated switching of a loop between superconducting state with different connectivity can be induced both by equilibrium noise (thermal fluctuation) and by an external non-equilibrium noise and it is difficult to distinguish the first and second influence. In order to state that only equilibrium noise induces the dc voltage observed in [27] one should be fully confident that the temperature of external noise in a wide frequency band does not exceed the temperature of measurement. It is very difficult to be sure even if because of the temperature difference inside and outside of the cryostat where the measurements any frequency regions. Because of \textit{T = 300 K} outside the cryostat the noise power can be close to the equilibrium one for the temperature \textit{T = 1.2 K} of measurement in [27] only for frequency regions which are strongly shielded or absorbed. The power for some frequency regions can be between $k_{B}1.2\Delta \omega $ and $k_{B}300\Delta \omega $ even without a external noise sources existing in our noisy world. 

But the claim that the dc power observed in [27] is induced by an external non-equilibrium electrical noise does not save the second law since already numerous observation of the persistent current at non-zero resistance are experimental evidence of its violation. This quantum phenomenon is enough long ago and well known. Most scientists state that the persistent current does not threaten the second law since it is equilibrium phenomenon and therefore no work can be extracted from the persistent current. Indeed, in the equilibrium state, in which the persistent current is observed, the free energy $F = E -- ST$ has minimum value and nobody can decrease a value below its minimum. But the internal energy $E$ can be decrease without any decrease of the free energy if the entropy $S$ decreases at the same time. Thus, this statement of defenders of the second law is turned into the statement that the second law can not be broken since it can not be broken. 

Since observation of any current $I$ at non-zero resistance $R > 0$ means the existence of energy dissipation with power $RI^{2}$ many scientists state that the persistent current is no quite current. But if it is correct why is the voltage proportional to this no-current is observed in [27]? In order to save the second law its defenders should explain this as well as they should give an explanation, alternative to the one proposed in [18], why the persistent current does not die down at $R > 0$. 

Now most scientist are fully confident that the second law can not be broken since as Arthur Eddington wrote in 1948 [36]: ``\textit{The second law of thermodynamics holds, I think, the supreme position among the laws of Nature. If someone points out to you that your pet theory of the universe is in disagreement with Maxwell's equations - then so much the worse for Maxwell's equations. If it is found to be contradicted by observation, well, these experimentalists do bungle things sometimes. But if your theory is found to be against the second law of thermodynamics I can give you no hope; there is nothing for it but collapse in deepest humiliation}''. Nevertheless enough many papers with challenge the absolute status of the second law were published in the last years [37-56]. More detailed Bibliography can be found at the web-site  http://www.sandiego.edu/secondlaw2002/\#Bibliography of First International Conference on Quantum Limits to the Second Law which was held July 29-31, 2002 in University of San Diego and at the web-site  http://www.ipmt-hpm.ac.ru/SecondLaw/. The Conference Proceedings were published by American Institute of Physics, http://proceedings.aip.org/proceedings/ confproceed/643.jsp.  

\bigskip

\noindent
[1] M. Smoluchowski, ``Gultigkeitsgrenzen des zweiten Hauptsatzes der Warmetheorie,'' in \textit{Vortrage uber kinetische Theorie der Materie und der Elektrizitat (Mathematische Vorlesungen an der Universitat Gottingen, VI)}. Leipzig und Berlin, B.G.Teubner, p.87 (1914).

\noindent
[2] Mario Gliozzi, Storia Della Fisica. Torino 1965. 

\noindent
[3] H. S. Leff and A. F. Rex, ``\textit{Maxwell's Demon: Entropy, Information, Computing}''. Princeton University Press, Princeton, 
1990.

\noindent
[4] J. C. Maxwell, \textit{Theory of Heat}, 4th ed., Longman's, Green, $\backslash $\& Co., London 
1985. 

\noindent
[5] A. Steane, ``Quantum Computing''. \textit{Rept.Prog.Phys.} \textbf{61}, 117 (1998). 

\noindent
[6] L.Szilard, \textit{Z.Phys.} \textbf{53}, 840 (1929) (English translation in ``\textit{Quantum Theory and Measurement}'', edited by J.A.Wheeler and W.H.Zurek, Princeton Univ.Press, Princeton, NJ, 1983). 

\noindent
[7] L. Brillouin, \textit{Science and Information}. Academic Press, New York, 1962. 

\noindent
[8] F.T.S. Yu, ``Entropy Information and Optics'', \textit{Optical Memory and Neural Networks} \textbf{9}, 75 (2000). 

\noindent
[9] W.H.Zurek, ``Thermodynamic Cost of Computation, Algorithmic Complexity 
and the Informational Metric'', \textit{Nature} \textbf{341}, 119 (1989).

\noindent
[10] C.M.Caves, ``Quantitative Limits on the Ability of a Maxwell Demon to 
Extract Work from Heat'', \textit{Phys.Rev.Lett.} \textbf{64}, 2111 (1990). 

\noindent
[11] Seth Lloyd, ``A quantum-mechanical Maxwell's demon''. quant-ph/9612034. 

\noindent
[12] W.H. Zurek, ``Algorithmic randomness, physical entropy, measurements, 
and the Demon of Choice''. quant-ph/9807007

\noindent
[13] V. Vedral, ``Landauer's erasure, error correction and entanglement''. quant-ph/9903049

\noindent
[14] C.H. Bennett and R. Landauer, ``The Fundamental Physical Limits of 
Computation'', \textit{Scientific American}, Jule p.38 (1985). 

\noindent
[15] C.H. Bennett, \textit{Int.J.Theor.Phys}. \textbf{21}, 905 (1982). 

\noindent
[16] A.E. Allahverdyan and Th.M. Nieuwenhuizen, ``Breakdown of Landauer bound for information erasure in the quantum regime'', \textit{Phys. Rev. E} \textbf{64}, 056117 (2001); cond-mat/0012284. 

\noindent
[17] R.P.Feynman, R.B.Leighton, and M.Sands, \textit{The Feynman Lectures on Physics}, vol.1, Addison-Wesley, 
Reading, Massachusetts, 1963. 

\noindent
[18] A.V. Nikulov, ``Quantum force in superconductor'' \textit{Phys. Rev. B} \textbf{64}, 012505 
(2001).

\noindent
[19] M.Tinkham, \textit{Introduction to Superconductivity}. McGraw-Hill Book Company (1975).

\noindent
[20] L.P.Levy, G.Dolan, J.Dunsmuir, and H.Bouchiat, \textit{Phys. Rev.Lett.} \textbf{64}, 2074 (1990).

\noindent
[21]V. Chandrasekhar, R.A.Webb, M.J.Brady, M.B.Ketchen, W.J.Gallagher, and A.Kleinsasser, ``Response of a Single, Isolated Gold Loop'' \textit{Phys. Rev.Lett.} \textbf{67}, 3578 (1991).

\noindent
[22] E.M.Q.Jariwala, P.Mohanty, M.B.Ketchen, and R.A.Webb, ``Diamagnetic Persistent Current in Diffusive Normal- Metal Rings'', \textit{Phys. Rev.Lett.} \textbf{86}, 1594 (2001).

\noindent
[23] W.A.Little and R.D.Parks, ``Observation of Quantum Periodicity in the Transition Temperature of a Superconducting Cylinder'' \textit{Phys.Rev.Lett.} \textbf{9}, 9 
(1962).

\noindent
[24] H.Vloeberghs, V.V.Moshchalkov, C. Van Haesendonck, R.Jonckheere, and 
Y.Bruynseraede, ``Anomalous Little-Parks Oscillations in Mesoscopic Loops'', 
\textit{Phys. Rev.Lett.} \textbf{69}, 1268 (1992).

\noindent
[25] H.Nyquist, \textit{Phys.Rev.} \textbf{32}, 110 (1928).

\noindent
[26] J.B.Johnson, \textit{Phys.Rev.} \textbf{32}, 97 (1928).

\noindent
[27] S.V.Dubonos, V.I.Kuznetsov, and A.V.Nikulov, "Segment of an 
Inhomogeneous Mesoscopic Loop as a DC Power Source" in Proceedings of 
10$^{{\rm t}{\rm h}}$ International Symposium "NANOSTRUCTURES: Physics and 
Technology" St Petersburg: Ioffe Institute, p. 350 (2002); 
http://xxx.lanl.gov/abs/physics/0105059. 

\noindent
[28] P.Santhanam, C.P.Umbach, and C.C.Chi, \textit{Phys.Rev. B} \textbf{40}, 11392 (1989); P.Santhanam et al. \textit{Phys.Rev.Lett.} \textbf{66}, 2254 (1991). 

\noindent
[29] A.V. Nikulov and I.N. Zhilyaev, \textit{J. Low Temp.Phys.} \textbf{112}, 227 (1998).

\noindent
[30] A.Barone and G.Paterno, \textit{Physics of the Josephson Effect}. Wiley, New York, 1982

\noindent
[31] A.V. Nikulov, ``One of Possible Applications of High-Tc Superconductors'', in \textit{SUPERMATERIALS}, edited by R.Cloots et al. Proceedings of the NATO 
ARW, Kluwer Academic Publishers, p. 183 (2000).

\noindent
[32] A.V. Nikulov, in Abstracts of XXII International Conference on Low Temperature Physics, Helsinki, Finland, p.498 (1999). 

\noindent
[33] A.V. Nikulov, ``A superconducting mesoscopic ring as direct-current generator'', \textit{Abstracts of NATO ASI "Quantum Mesoscopic Phenomena and Mesoscopic Devices in Microelectronics"}  Ankara, Turkey, p.105 (1999)

\noindent
[34] A.V. Nikulov, ``A system of mesoscopic superconducting rings as a microrefrigerator''. \textit{Proceedings of the Symposium on Micro- and Nanocryogenics}, Jyvaskyla, Finland, p.68 (1999).

\noindent
[35] V.V.Aristov and A.V.Nikulov, \textit{Abstracts of Fourth APAM Topical Seminar}, Seoul, Korea, p.25 (2000).

\noindent
[36] A.S. Eddington, \textit{The Nature of the Physical World. Macmillan}, New York, 1948. 

\noindent
[37] Th. M. Nieuwenhuizen and A. E. Allahverdyan, ``Extraction of work form 
a single thermal bath in quantum regime''. \textit{Phys. Rev. Lett.} \textbf{85,} 1799 (2000).

\noindent
[38] Th. M. Nieuwenhuizen and A. E. Allahverdyan, ``Statistical thermodynamics of quantum Brownian motion: Construction of perpetuum mobile 
of the second kind''. \textit{Phys. Rev. E} \textbf{66}, 036102 (2002)

\noindent
[39] V. Capek and D.P.Sheehan, ``Quantum mechanical model of a plasma system; a challenge to the second law of thermodynamics''. \textit{Physica A} \textbf{304}, 461 (2002). 

\noindent
[40] V. Capek and J.Bok, ``A thought construction of working perpetuum 
mobile of the second kind''. \textit{Czeck J. Phys.} \textbf{49} 1645 (1999).

\noindent
[41] V. Capek, ``Twilight of a dogma of statistical thermodynamics''. \textit{Mol. Cryst. Liq. Cryst.} \textbf{355}, 13 (2001). 

\noindent
[42] V. Capek and J. Bok, ``Violation of the second law of thermodynamic in 
the quantum microworld''. \textit{Physica A} \textbf{290} 379 (2001)

\noindent
[43] V. Capek, ``Zeroth and second laws of thermodynamics simultaneously questioned in the quantum microworld''. \textit{European Physical Journal B} \textbf{25}, 101 (2002) (see also 
at http://arxiv.org/abs/cond-mat/0012056). 

\noindent
[44] L.G.M.Gordon, ``Maxwell's demon and detailed balancing''. \textit{Found. Phys.} \textbf{13}, 
989 (1983). 

\noindent
[44] L.G.M.Gordon, ``Brownian movement and microscopic irreversibility''. \textit{Found. Phys.} \textbf{11}, 103 (1981). 

\noindent
[46] R.L.Liboff, ``Maxwell's demon and the second law of thermodynamics''. \textit{Found. Phys. Lett.} \textbf{10}, 89 (1997).

\noindent
[47] D.P.Sheehan, ``The second law and chemically-induced, steady-state pressure gradients: controversy, corroboration and caveats'', \textit{Phys. Lett. A} \textbf{280}, 
185 (2001). 

\noindent
[48] D.P.Sheehan, J. Glick, J.D. Means, ``Steady-state work by an 
asymmetrically inelastic gravitator in a gas: a second law paradox'', 
\textit{Foundations of Physics} \textbf{30}, 1227 (2000). 

\noindent
[49] D.P.Sheehan, J. Glick, ``Gravitationally-induced, dynamically-maintained, steady-state pressure gradients'', \textit{Phys. Script.} \textbf{61}, 635 
(2000). 

\noindent
[50] D.P.Sheehan, Reply to ``Comment on `Dynamically maintained steady-state 
pressure gradients.''' \textit{Phys. Rev. E} \textbf{61} 4662 (2000).

\noindent
[51] D.P.Sheehan, ``Four paradoxes involving the second law of thermodynamics''. \textit{J. Sci. Explor.} \textbf{12}, 303 (1998).

\noindent
[52] D.P.Sheehan, ``Dynamically-maintained, steady-state pressure 
gradients''. \textit{Phys. Rev.} $E$ \textbf{57}, 6660 (1998).

\noindent
[53] D.P.Sheehan, ``A paradox involving the second law of thermodynamics''. 
\textit{Phys. Plasmas} \textbf{2}, 1893 (1995).

\noindent
[54] G.M. Zaslavsky and M.Edelman, ``Maxwell's demon as a dynamical model''. 
\textit{Phys. Rev. E} \textbf{56}, 5310 (1997). 

\noindent
[55] G.M. Zaslavsky, ``From Hamiltonian chaos to Maxwell's demon''. \textit{Chaos} 
\textbf{5,} 653 (1995).

\noindent
[56] P. Weiss, ``Breaking the Law: Can quantum mechanics + thermodynamics =perpetual motion?" \textit{Science News}, \textbf{158}, 234. (2000); http://www.sciencenews.org/ 20001007/toc.asp

\end{document}